\documentclass[12pt]{article}
\usepackage{amsmath}
\usepackage{amssymb,amsthm}
\usepackage{bbold}
\usepackage{color}

\title{Irreversibility, the time arrow and a dynamical proof of the second law of thermodynamics}

\author{Walter F. Wreszinski\footnote{wreszins@gmail.com, 
Instituto de Fisica, Universidade de S\~ao Paulo (USP), Brazil}}        

\begin{document}

\maketitle

\begin{abstract}
We provide a dynamical proof of the second law of thermodynamics, along the lines of an argument of Penrose and Gibbs, making crucial use of the upper semicontinuity of the mean entropy proved by Robinson and Ruelle and Lanford and Robinson. An example is provided by a class of models of quantum spin systems introduced by Emch and Radin. Consequences regarding irreversibility and the time arrow, as well as possible extensions to quantum continuous systems, are discussed. 
\end{abstract}

\section{Introduction: definition of an adiabatic transformation}

In an important series of papers, Lieb and Yngvason proposed an axiomatic foundation of the second law of thermodynamics (\cite{LYPR}, \cite{LYNE}). In \cite{LYNE}, pg. 3, they remark that the additivity property for the entropy ''is one of the most difficult to try to prove if ever there is a mathematical proof of the second law from assumptions about dynamics''.

In this paper we attempt to provide such a dynamical proof, along the lines of an argument due to Gibbs and more precisely formulated by Penrose (see \cite{Pen}, pg.1959, see also his book \cite{Pen1}), using the natural concept of state as a normalized linear functional over an algebra of observables \cite{BRo1},\cite{BRo2}, which is valid for both classical and quantum statistical mechanics, in the former case the algebra being Abelian. We shall argue that the relevant quantity to be considered in a dynamical treatment is the mean entropy, and, indeed, in the case of classical statistical mechanics, Ruelle has conjectured that, concerning the problems of evolution ((iv), pg. 1666, of \cite{RuJMP}), ''an equilibrium state would be a fixed point for the evolution of states. It is unclear to the author whether the evolution of an infinite system should increase its entropy per unit volume. Another possibility is that, when the time tends to $\infty$, a state has a limit with strictly larger entropy.''

We show in theorem 3.1 that all three remarks and conjectures above take place for a class of models of quantum spin systems introduced by Emch \cite{Em1} and Radin \cite{Ra}. The general argument is applicable to both classical and quantum statistical mechanics, but is presented in section 3 in the framework of quantum spin systems, for technical reasons. The main property of the mean entropy which is used - is valid both in the classical case, as shown by Robinson and Ruelle \cite{RR}, and in the quantum case, as proved by Lanford and Robinson \cite{LanRo}, but the treatment of classical evolution requires consideration of the momenta, omitted in \cite{RR}. Ruelle's suggestion of using the one-point compactification of $\mathbf{R}^{\nu}$ to treat the momenta (\cite{RuJMP}, pg 1666) might work, but the technical details remain open.

The reason for considering the mean entropy rather than the full entropy constructed in (\cite{LYPR},\cite{LYNE}) in the dynamical situation is, of course,  that the limit $t \to \infty$ may not, and in general will not, commute with the limit of infinite volume. Naturally, in practice it will not be necessary to pass to infinite systems and times, but we must be sure that the error in the approximations may be made arbitrarily small for sufficiently large volume and time, which is not possible if, for instance, the rate of approach to equilibrium grows indefinitely, when the volume tends to infinity. The consideration of infinite systems is even more imperative in the case of quantum systems, since for finite systems the Hamiltonian has a point spectrum and quasiperiodic observables (see section 1.5 of \cite{Pen}).

The main result is theorem 3.1. Its implication regarding irreversibility is discussed in remark 3.1, and the time-arrow problem is left to section 5. 
The problem of irreversibility has been studied intensively in recent years by J. Froehlich and several collaborators, a review with references being \cite{Fro}, as well as by Haag \cite{Haag}; the issue of the role of the observer in these studies is briefly mentioned in connection with our approach at the conclusion. General important references connected with the present paper, besides (\cite{Pen}, \cite{Pen1}, \cite{LP}), are the recent \cite{GLTZ} and \cite{NarWre}. Concerning the second law, the paper by Abou-Salem and Fr\"{o}hlich \cite{AF} is a complementary approach to the present paper: see also the additional references given there.

We end this introduction by briefly describing our approach, which is clearer in the classical case, following closely section 2 of \cite{Pen}. The second law of thermodynamics asserts that there exists an additive thermodynamic function - the entropy - which can increase but not decrease in an adiabatic process, that is, a process in which no heat enters or leaves the system. The latter is modelled using a Hamiltonian $H(t)$ which depends on time. The (classical) state of the system being described by a phase space density $\rho(x)$ (defined for almost every $x \in \Gamma$, where $\Gamma$ denotes phase space, in the special case where the measures in \cite{RR} are absolutely continuous with respect to Lebesgue measure), the adiabatic process may be assumed to start at time $t=0$, when the Hamiltonian is $H(t=0)$, and end at time $t=1$, when the Hamiltonian has changed to $H(t=1)$, and remain constant thereafter, further assumed to yield an ergodic and mixing motion. The final equilibrium is described, however, not by this time-dependent phase-space density, but by a new phase-space density $\bar{\rho}$, the coarse-grained or weak limit of the phase space density, which, in the \textbf{classical} case (by the Birkhoff ergodic theorem) may be calculated by the time-averaging prescription
$$
\bar{\rho}(x) = \lim_{T\to \infty} \frac{\int_{1}^{1+T} \rho_{t}(x)dt}{T}
\eqno{(1.1)}
$$
where
$$
\rho_{t}(x) = \rho_{0}(U_{-t}(x))
\eqno{(1.2)}
$$
by Liouville's theorem and causality (see (1.9), pg. 1942 of \cite{Pen}), with $x_{t}=U_{t}(x_{0})$ denoting the evolution of an initial point of phase space by Hamiltonian dynamics. Define, now, the entropy $S(\rho)$ associated to the state defined by the phase space density $\rho$, by (compare also with (2.6) of \cite{RR}) 
$$
S(\rho) \equiv   -\int_{\Gamma} \rho(x) \log(\rho(x)) dx
\eqno{(1.3)}
$$
Using, now, Liouville's theorem and the concavity of the functional $\rho \to S(\rho)$, an argument attributed by Penrose to Gibbs (\cite{Pen},pg. 1959) -which we call the Penrose-Gibbs argument - yields
$$
S(\rho_{0}) \le S(\bar{\rho})
\eqno{(1.4)}
$$
Although this argument is certainly correct, we believe that the equality sign will necessarily occur in (1.4) if a finite region is considered, since then the functional (1.3) is continuous (in the natural weak-* topology, see \cite{RR}).

For classical systems, it is the mixing property which determines the approach to equilibrium in the sense
$$
\lim_{t\to \infty} \int_{\Gamma} \rho_{t}(x) G(x) dx = \langle G \rangle_{eq} \equiv \int_{\Gamma} G(x) \bar{\rho}(x) dx
\eqno{(1.5)}
$$
for $G$ an observable (a continuous integrable function over phase space), see \cite{Pen}, (1.35), pg. 1949 and the references given there and, in addition, \cite{Sinai1}. That is, $\rho_{t}$ converges pointwise, as a state over the classical observable algebra \cite{RR}: this implies convergence of the ergodic averages on the r.h.s to the same limit, but, of course, not conversely. (1.5) means that mixing systems are ''memoryless'', i.e., they posess a stochastic character which justifies equilibrium statistical mechanics. This approach is well explained by Lebowitz and Penrose \cite{LP}, and the microscopic mechanism of the loss of memory  is believed to occur also in quantum continuous systems, see section 6.

A state $\rho$ will be defined as a  normalized, positive linear functional over the algebra of observables ${\cal A}$ of an infinite system (which may be classical or quantum, as previously described). Readers unfamiliar with this notion may consult the classic book \cite{Sewell1}, or picture it as a thermodynamic limit of ground states $\rho(A)= \lim_{\Lambda \to \mathbf{Z}^{\nu}} (\Omega_{\Lambda}, A \Omega_{\Lambda})$ or Gibbs states 
$\rho(A) = \frac{tr(\exp(-\beta H_{\Lambda})A)}{tr(\exp(-\beta H_{\Lambda})}$, in the lattice case, where $\nu$ is the space dimension and $\Omega_{\Lambda}$ is the ground state of the Hamiltonian $H_{\Lambda}$ and $\beta$ is the inverse temperature.

For quantum spin systems, the algebra of observables is the algebra generated by the Pauli spin operators at each site, together with the identity. An automorphism $\alpha_{t}(A)$ of the algebra may be viewed as the limit in the operator norm of the time evolutes $\exp(iH_{\Lambda}t) A \exp(-iH_{\Lambda}t)$: one speaks of a $C*$ dynamical system $({\cal A},\alpha_{t})$. 

We refer to \cite{BRo1}\cite{BRo2} for the description of the set of states $\rho$ over the C* algebra ${\cal A}$ of quasi-local observables describing an infinite (quantum spin) system. We wish to describe systems with only a finite number of spins in each bounded region $\Lambda \subset \mathbf{Z}^{\nu}$ , so that to ${\cal A}$ is associated a collection $\{{\cal A}(\Lambda)\}$ of C* subalgebras of  ${\cal A}$ corresponding to finite subsets $\Lambda$ of $\mathbf{Z}^{\nu}$, satisfying certain conditions, summarized in section 2 of \cite{LanRo}. 

The above discussion suggests the following definition:

\textbf{Definition 1.1}

Let a $C*$ -dynamical system $({\cal A},\alpha_{t}$ be given. An \textbf{adiabatic transformation} consists of two successive steps. The first step, called preparation of the state, starts at some $t=-r$, with $r>0$, when the state $\rho_{-r}$ is invariant under $\alpha_{-r}$ and the hamiltonian is $H_{-r}$, and ends at $t=0$, when the Hamiltonian is $H_{0}$, such that
$$
H_{-r} = H_{0}
\eqno{(1.6)}
$$
and the state is $\rho_{0}=\rho$. The second step is a dynamical evolution of the state of the form
$$
t \in \mathbf{R} \to \rho_{t} \equiv \rho \circ \alpha_{t}
\eqno{(1.7)}
$$

\textbf{Remark 1.1}

Definition 1.1 is nontrivial only if the preparation of the state yields $\rho \circ \alpha_{t} \ne \rho$, i.e., if the initial state $\rho$ is not invariant under the automorphism $\alpha_{t}$.

\textbf{Remark 1.2}

In \cite{LYPR}, Lieb and Yngvason discuss ''heat'', which they dismiss with the harsh words: ''No one has ever seen heat, nor will it ever be seen, smelled or touched''. Note that definition 1.1 avoids this concept. Another important point is that the word ''adiabatic'' in definition 1.1 does not imply ''slow'' or ''quasi-static'', similarly to the definition of adiabatic accessibility in \cite{LYPR} (see the discussion in \cite{LYPR}, page 17, Remark). Indeed, we shall see in the examples of section 4 that the initial state need not be a ''gentle'' perturbation of the equilibrium state. The concept of adiabatic transformation in classical mechanics (\cite{Thirr3rd}, Cor. 3.5.19, pg. 153) is therefore unrelated to definition 1.1 in the classical case.  

We now describe the framework, as well as the class of models, in greater detail.

\section{The framework and two propositions for the specific entropy}

We assume that the initial state $\rho$ (i.e., at time $t=0$) is assumed to define a family $\{\rho_{\Lambda}\}$ of density matrices, each acting on a Hilbert space
$$
{\cal H}_{\Lambda} = \otimes_{x \in \Lambda} {\cal H}_{x}
\eqno{(2.1)}
$$
with
$$
\dim {\cal H}_{x} = D
\eqno{(2.2)}
$$
and further satisfying the conditions of compatibility, translation invariance (see section 2 of \cite{LanRo}) and normalization, i.e,
$$
Tr_{{\cal H}_{\Lambda}} (\rho_{\Lambda}) = 1
\eqno{(2.3)}
$$

On ${\cal H}_{\Lambda}$ we assume we are given a self-adjoint Hamiltonian $H_{\Lambda}$, which defines the dynamics by the automorphisms
$$
\alpha_{t}(A) = \mbox{ norm } \lim (\exp(i H_{\Lambda}t) A \exp(-i H_{\Lambda}t)) \mbox{ as } \Lambda \nearrow \mathbf{Z}^{\nu} 
\eqno{(2.4)}
$$
The above limit holds for $A \in {\cal A}_{L}$ initially, where ${\cal A}_{L}$ denotes the strictly local algebra, and then 
the extension of the above limit to $A \in {\cal A}$ defines a group of automorphisms $t \to \alpha_{t}$ of ${\cal A}$, for a 
large class of quantum spin systems \cite{Rob}.

For each state $\rho$, we define a family of entropy functions (or entropies) $S(\rho_{\Lambda})$ by
$$
S(\rho_{\Lambda}) = -Tr_{{\cal H}_{\Lambda}}(\rho_{\Lambda} \log(\rho_{\Lambda}))
\eqno{(2.5)}
$$
Let us define, for $(a_{i}, \cdots a_{\nu}) \in \mathbf{Z^{\nu}} \mbox{ with } a_{1}>0, \cdots, a_{\nu}>0$, the parallelepiped $\Lambda(a)=\{x \in \mathbf{Z^{\nu}}; 0<x_{i}<a_{i}, \mbox{ for } i=1, \cdots, \nu$ with volume $V(\Lambda(a))=\prod_{i=1}^{\nu} a_{i}$. We have:

\textbf{Theorem 2.1} The entropy function satisfies the inequalities
$$
0 \le S(\rho_{\Lambda}) \le V(\Lambda(a)) \log D
\eqno{(2.6)}
$$
and is subadditive, i.e., 
$$
S(\rho_{\Lambda_{1}\cup \Lambda_{2}}) \le S(\rho_{\Lambda_{1}}) + S(\rho_{\Lambda_{2}}) \mbox{ if } \Lambda_{1}\cap \Lambda_{2}=\phi
\eqno{(2.7)}
$$
If the family $\rho=\{\rho_{\Lambda}\}$ is appropriate, i.e., satisfies the conditions of compatibility, translation invariance and normalization, then: the \textbf{mean entropy} $s(\rho)$ satisfies: 
$$
s(\rho) \equiv \lim_{a_{1}, \cdots a_{\nu} \to \infty} \frac{S(\rho_{\Lambda(a)})}{V(\Lambda(a))}
\eqno{(2.8)}
$$
exists, and
$$
s(\rho) = \inf_{a_{1}, \cdots a_{\nu}} \frac{S(\rho_{\Lambda(a)})}{V(\Lambda(a))}
\eqno{(2.9)}
$$
  
For the proof of theorem 2.1, see theorems 1 and 2 of \cite{LanRo}.

For each finite subset $\Lambda$ of $\mathbf{Z}^{\nu}$, and for each time $t \ge 0$, we define states $\rho_{\Lambda,t}$ on ${\cal A}_{\Lambda}$ by
$$
\rho_{\Lambda,t}(A) = tr_{{\cal H}_{\Lambda}}(\rho_{\Lambda}(t) A) \mbox{ for } A \in {\cal A}_{\Lambda}  
\eqno{(2.10.1)}
$$
where
$$
\rho_{\Lambda}(t) \equiv \exp(-i H_{\Lambda}t) \rho_{\Lambda} \exp(i H_{\Lambda}t)
\eqno{(2.10.2)}
$$
Let $(\omega)_{\Lambda}$ denote the restriction of a state $\omega$ on the quasi-local algebra ${\cal A}_{\Lambda}$, and $A/B$ denote
the complement of $B$ in $A$. We have the following crucial estimate:

\textbf{Proposition 2.1}

Let $\Lambda$ be a fixed finite subset of $\mathbf{Z^{\nu}}$. The following inequality holds:

\begin{eqnarray*}
|S((\rho_{t})_{\Lambda} - S((\rho_{\Lambda,t})| \le\\
\le |\Lambda_{0}| \log D \sup_{A\in{\cal A}_{\Lambda_{0}},||A||=1} ||\alpha_{t,\Lambda,A}||+\\
+ \log 2 + 2|\Lambda / \Lambda_{0}| \log D
\end{eqnarray*}$$\eqno{(2.11.1)}$$
where $\rho_{t}$ is defined by (1.7), $\Lambda_{0} \subset \Lambda$, and
$$
\alpha_{t,\Lambda,A} \equiv \alpha_{t}(A)-\exp(itH_{\Lambda})A\exp(-itH_{\Lambda})
\eqno{(2.11.2)}$$

\textbf{Proof}

Using the subadditivity property of the entropy, we have
\begin{eqnarray*}
|S((\rho_{t})_{\Lambda} - S((\rho_{\Lambda,t})| \le \\
\le |S((\rho_{t})_{\Lambda_{0}} - S((\rho_{\Lambda,t})_{\Lambda_{0}})|+\\
+ 2|\Lambda / \Lambda_{0}| \log D
\end{eqnarray*}$$\eqno{(2.12)}$$
By Fannes' inequality \cite{Fannes} we have
\begin{eqnarray*}
|S((\rho_{t})_{\Lambda_{0}} - S((\rho_{\Lambda,t})_{\Lambda_{0}})| \le \\
\le |\Lambda_{0}|\log D ||(\rho_{t})_{\Lambda_{0}} -(\rho_{\Lambda,t})_{\Lambda_{0}}||_{1} + \log 2
\end{eqnarray*}$$\eqno{(2.13)}$$
where $||\cdot||_{1}$ denotes the trace norm. We now estimate the trace norm difference:

\begin{eqnarray*}
||(\rho_{t})_{\Lambda_{0}} -(\rho_{\Lambda,t})_{\Lambda_{0}}||_{1} =\\
= \sup_{A\in {\cal A}_{\Lambda_{0}},||A||=1} tr_{{\cal H}_{\Lambda_{0}}}([(\rho_{t})_{\Lambda_{0}} -(\rho_{\Lambda,t})_{\Lambda_{0}}]A) = \\
= \sup_{A\in {\cal A}_{\Lambda_{0}},||A||=1} tr_{{\cal H}_{\Lambda}}[\rho_{\Lambda}(\lim_{n} \sigma_{n,\Lambda,A} - \\
-\exp(itH_{\Lambda})A \exp(-itH_{\Lambda}))]
\end{eqnarray*}$$\eqno{(2.14)}$$
where
\begin{eqnarray*}
\sigma_{n,\Lambda,A} \equiv \\
tr_{{\cal H}_{\Lambda_{n} / \Lambda}}\rho_{\Lambda_{n} / \Lambda}\exp(itH_{\Lambda_{n}})A \exp(-itH_{\Lambda_{n}})
\end{eqnarray*}$$\eqno{(2.15)}$$

(2.15) implies
\begin{eqnarray*} 
||(\rho_{t})_{\Lambda_{0}} -(\rho_{\Lambda,t})_{\Lambda_{0}}||_{1} \le \\
\le \sup_{A\in {\cal A}_{\Lambda_{0}},||A||=1}||\alpha_{t,\Lambda,A}||
\end{eqnarray*}$$\eqno{(2.16)}$$
with $\alpha_{t,\Lambda,A}$ defined by (2.11.2). q.e.d. 

We now specify our quantum spin models in greater detail. 

Let, as in \cite{Nachtergaele}, appendix, $\Gamma,d$ denote the countable metric space $\mathbf{Z}^{\nu}$ with the metric 
$d$ given by the $l^{1}$ - metric, with $d(x,y)=|x-y| \equiv \sum_{j=1}^{\nu}|x_{j}-y_{j}|$. In (2.17), $\Phi(X)$ will denote,
for simplicity, a two-body interaction potential, i.e., $X= (x,y)$, with $x,y \in \mathbf{Z}^{\nu}$: extensions are possible
but would require extra concepts and notations along the lines of \cite{Nachtergaele}.
Let the Hamiltonian for a finite region $\Lambda$ be defined by 

\begin{eqnarray*}
H_{\Lambda} \equiv \sum_{X \subset \Lambda} \Phi(X)\\
\mbox{ where } \Phi(X) = \Phi(X)^{*} \mbox{ and } \\
||\Phi(X)|| \le \mbox{ const. } \exp(-\alpha d(x,y))\\
\mbox{ for } \alpha > 0 \mbox{ and } X=(x,y), x,y \in \mathbf{Z}^{\nu}\\
\end{eqnarray*}$$\eqno{(2.17)}$$
Let $F$ denote a non-increasing
function $F:[0,\infty) \to [0,\infty)$ which is uniformly integrable and satisfies the conditions in \cite{Nachtergaele}, appendix.
Let $F_{g}(r) = \exp(-g(r)) F(r)$ with $g$ a non-negative, non-decreasing, subadditive on $[0,\infty)$, and $F$ of finite
norm, as defined by $||F|| \equiv \sup_{\mathbf{Z}^{\nu}} \sum_{x \in \mathbf{Z}^{\nu}}F(d(x,y))$. One has, then ,the bound (\cite{Nachtergaele}, A21)):
$$
\sum_{x \in X}\sum_{y \in Y} F_{g}(d(x,y)) \le |X|||F|| \exp(-g(d(x,y))
\eqno{(2.18)}
$$
 
The next proposition uses a theorem on the comparison of dynamics (Theorem 3.4 of \cite{Nachtergaele}), which is connected with
bounds first proved by Lieb and Robinson \cite{LiR}. The review \cite{Nachtergaele} contains a comprehensive discussion
of several related issues. 

\textbf{Proposition 2.2}

Let $\rho$ be a translation invariant or periodic product state on ${\cal A}$, the quasi-local algebra of a quantum spin system
on $\mathbf{Z}^{\nu}$, with a dynamics $\{\alpha_{t}\}_{t \in \mathbf{R}}$ generated by a translation-invariant family of 
Hamiltonians $H_{\Lambda_{n}}, n \ge 1$ given by (2.17). Then, the specific entropy of $\rho_{t}=\rho \circ \alpha_{t}$, denoted
by $s(\rho_{t})$, satisfies

$$
s(\rho_{t}) = \lim_{n} \frac{1}{|\Lambda_{n}|}S(\exp(-itH_{\Lambda_{n}})\rho_{\Lambda_{n}}\exp(itH_{\Lambda_{n}}))
\eqno{(2.19)}
$$
where $\Lambda_{n}$ is a sequence of boxes tending to $\mathbf{Z}^{\nu}$.

\textbf{Proof} 

By definition 
$$
s(\rho_{t}) = \lim_{n} \frac{1}{|\Lambda_{n}|} S((\rho_{t})_{\Lambda_{n}})
\eqno{(2.20)}
$$
By (2.17), (3.69) of Theorem 3.4 of \cite{Nachtergaele}, and choosing $g(r)= \beta r$ with $\beta$ 
any constant such that $0 < \beta < \alpha$, we find

\begin{eqnarray*}
||\alpha_{t}(A)-\exp(itH_{\Lambda})A\exp(-itH_{\Lambda})|| \le\\
\le ||A|| C_{\alpha}(t) \exp(-\beta d(\Lambda_{0},\mathbf{Z}^{\nu}\ \Lambda_{0}))
\end{eqnarray*}$$\eqno{(2.21)}$$
where $C_{\alpha}(t)$ is a constant. We now pick, for each $n \ge 1$, 
$\Lambda_{n,0} \subset \Lambda_{n}$ such that: i) $\frac{|\Lambda_{n,0}|}{|\Lambda_{n}|} \to 1$; 
ii.) $\frac{|\Lambda_{n} / \Lambda_{n,0}|}{|\Lambda_{n}|} \to 0$; iii.) $d(\Lambda_{n,0}, \mathbf{Z}^{\nu} / \Lambda_{n}) \to \infty$, 
thereby obtaining from (2.21) and proposition 2.1:
$$
\lim_{n} \frac{1}{|\Lambda_{n}|}|S((\rho_{t})_{\Lambda_{n}})-S(\rho_{\Lambda_{n},t})| = 0
\eqno{(2.22)}
$$
Since $s(\rho_{t})$ exists as the limit (2.19), this completes the proof. q.e.d.

\textbf{Remark 2.1}

The necessity of including exponentially decreasing interactions, but of infinite range, in the above proposition is due to the fact that 
for finite-range interactions, $(\rho \circ \alpha_{t})(A)$ is, for any $A \in {\cal A}$, in general an almost periodic function
of time. Indeed, this was proved 
by Radin \cite{Ra} for the generalized Ising model (GIM) in his lemma 2, pg. 2951. There is, therefore, no hope of proving the forthcoming 
hypothesis (3.1) of the existence of a ''coarse-grained state'' for such interactions.
Since finite range interactions are approximations to the short range interactions of infinite range occurring in nature, 
of which (2.17) are an example, this does not pose any problem.
In section 5 we illustrate that the limit on the r.h.s. of (3.1) depends \textbf{both} on the infinite volume limit 
and on the fact that the interactions are of infinite range.

\section{A dynamical proof of the second law for quantum spin systems}

In this section we describe a dynamical proof of the second law for quantum spin systems. For clarity, we assume that $\rho_{0}$ is already given as the state at the beginning of the second step of the adiabatic transformation, and come back to the preparation of the state in section 5. The arguments are valid without essential changes for fermion systems and quantum lattice systems, as treated by Lanford and Robinson in \cite{LanRo}.

Let $\rho$ satisfy the assumptions of proposition 2.2 and define the \textbf{coarse-grained state} by the following limit, 
provided it exists in the weak* topology:
$$
\bar{\rho} \equiv \lim_{t \to \infty} \rho_{t}
\eqno{(3.1)}
$$

We have

\textbf{Theorem 3.1}
$$
s(\rho_{0}) \le s(\bar{\rho})
\eqno{(3.2)}
$$
In words: the mean entropy of the initial state may increase, but cannot decrease, towards that of the coarse-grained state (3.1), under an adiabatic transformation (that is, with $\rho_{t}$ in (3.1) defined as in definition 1.1).

\textbf{Proof} By (2.9) and the continuity of $S$ in the weak* topology, it follows that $s$ is upper semicontinuous in the weak* topology (see theorem 3 of \cite{LanRo}), from which it follows that
$$
\limsup_{ t\to \infty} s(\rho_{t}) \le s(\lim_{t \to \infty} {\rho}_{t})= s(\bar{\rho})
\eqno{(3.3)}
$$
By (2.19) of proposition 2.2 and the invariance of the trace under unitary transformations, it follows that $ s(\rho_{t}) = s(\rho_{0})$ which, 
together with (3.3), yields (3.1). q.e.d.

\textbf{Remark 3.1} Theorem 3.1 implies \textbf{irreversibility} in the following sense: if $s(\rho_{0}) < s(\bar{\rho})$, the reversed evolution from $t= \infty$ to $t=0$ does not take place. The meaning of $t= \infty$ is: $t$ is a ''macroscopic'' time, i.e., much greater than the (microscopic) characteristic times of the system. This is in agreement with the infinite volume limit taken in the definition of the specific entropy, which makes it a macroscopic quantity. For the experiments of free induction decays in solids described in \cite{LowNo}, relevant to the models described in section 4, such characteristic times are of the order of microseconds, for atomic resonances described in \cite{Wre}), they are of the order of $10^{-8}$ seconds. Although the monotonicity property for the specific entropy $s(\rho_{t}) \le s(\rho_{r}) \mbox{ if } t \le r$ was not established in theorem 3.1, it may be argued that, by the same arguments, this monotonicity is also expected only for $t,r$ macroscopic, and, if $t < r$, $t$ and $r$ are ''infinitely apart'' in comparison with microscopic times. Theorem 3.1 may, therefore, be adequate as a statement of irreversibility. 

\textbf{Remark 3.2} We have assumed, in remark 3.1, that a given time direction is given. Thus, the irreversibility problem is distinct from the \textbf{time arrow} problem, to which we come back in section 5, and is related to the preparation of the state in definition 1.1. 

Theorem 3.1 shows the possibility of strict inequality in (3.2) due to the upper semicontinuity of $s$. We now show that this possibility indeed occurs for a class of quantum spin systems.

\section{Application to the generalized Ising model}

We now consider the example of the generalised Ising model (GIM) introduced by Emch \cite{Em1} and Radin \cite{Ra}, where, in (2.17),

$$
H_{\Lambda} = \frac{1}{2} \sum_{j,k \in \Lambda \times \Lambda} \Phi(j,k)
\eqno{(4.1.1)}
$$
where
$$
\Phi(j,k) \equiv \epsilon(|j-k|) \sigma_{z}^{j}\sigma_{z}^{k}
\eqno{(4.1.2)}
$$
and
$$
\sum_{j \in \mathbf{Z}^{\nu}} \epsilon(|j|) < \infty
\eqno{(4.1.3)}
$$

This \textbf{Generalized Ising Model} (GIM) was shown to describe a non-Markovian approach to equilibrium consistent with some experiments on pulsed nuclear magnetic resonance of nuclei in rigid lattices \cite{LowNo} in the case $\nu = 1$ and in the forthcoming special case of the exponential model with $\xi=2$. In these models, a calcium fluorine crystal is placed in a magnetic field, thus determining the z-direction, and allowed to reach thermal equilibrium, and a rf pulse is applied which turns the net nuclear magnetization in the x direction.

The model's rich asymptotic behavior seems closer to what is expected from a system of interacting spins, in contrast to the XY model, which seems closer to a free system \cite{Ro}. 

In the special case (4.1), the automorphisms (2.4) may be obtained explicitly. Let
$$
\mathbf{Z}^{\nu} \div V \equiv \mathbf{Z}^{\nu} \times \mathbf{Z}^{\nu}/((\mathbf{Z}^{\nu}/V) \times ((\mathbf{Z}^{\nu}/V) 
\eqno{(4.2)}
$$
We have (proposition 1 of \cite{Ra})
\begin{eqnarray*}
A \in {\cal A}_{\Lambda} \to \alpha_{t}(A) = \exp(i\tilde{H_{\Lambda}}t) A \exp(-i\tilde{H_{\Lambda}}t)\\
\mbox{ where } \tilde{H_{\Lambda}}=\frac{1}{2}\sum_{(j,k)\in (\mathbf{Z}^{\nu} \div \Lambda)} \Phi(j,k)
\end{eqnarray*}$$\eqno{(4.3)}$$
and $\Phi(j,k)$ was defined in (4.1.2). We now compare $\alpha_{t}(A)$ with $\alpha_{t}^{\Lambda}(A) \equiv \exp(iH_{\Lambda}t)A \exp((-iH_{\Lambda}t)$.
By proposition 1 of \cite{Ra},
\begin{eqnarray*}
\alpha_{t}(\sigma_{x}^{0})= \sigma_{x}^{0}\cos(2t\sum_{j\in \mathbf{Z}^{\nu}} R_{j})\\
-\sigma_{y}^{0}\sin(2t\sum_{j\in \mathbf{Z}^{\nu}} R_{j})
\end{eqnarray*}
$$\eqno{(4.4.1)}$$
while
\begin{eqnarray*}
\alpha_{t}^{\Lambda}(\sigma_{x}^{0}) = \sigma_{x}^{0}\cos(2t\sum_{j\in \Lambda} R_{j})-\\
-\sigma_{y}^{0}\sin(2t\sum_{j\in \Lambda} R_{j})
\end{eqnarray*}
$$\eqno{(4.4.2)}$$
where
$$
R_{j} = \epsilon(|j|)\sigma_{z}^{j}
\eqno{(4.5)}
$$ 

We now apply (4.4), (4.5) to some special states. let, now, $\nu =1$, and consider the state 
$\rho_{0}= \otimes_{j\in \mathbf{Z}} P_{j}^{+,x} \mbox{ where }\sigma_{x}^{j} P_{j}^{+,x} = P_{j}^{+,x}$, and its corresponding finite volume
version $\rho_{\Lambda,0}$; consider $\rho_{\Lambda,t}$ defined in (2.10). By (4.4.2),
$$
\rho_{t,\Lambda}(\sigma_{x}^{0}) = \prod_{j \in \Lambda} \cos(2\epsilon(|j|)t)
\eqno{(4.6)}
$$
Under condition (4.1.3),
\begin{eqnarray*}
\lim_{\Lambda \nearrow \mathbf{Z}} \cos(2\epsilon(|j|)t) = P\\
\mbox{ where } P \equiv \prod_{j\in \mathbf{Z}} \cos(2\epsilon(|j|)t)
\end{eqnarray*}$$\eqno{(4.7)}$$

The infinite product $P$ exists and equals zero for a variety of interactions. For instance, if $\epsilon(|j|)= 2^{-|j|}$, Vieta's identity
(see \cite{Em1}) yields $P = (\frac{\sin t}{t})^{2}$. Thus 
$\lim \rho_{t}(\sigma_{x}^{0})= 0 = \bar{\rho}(\sigma_{x}^{0})= (\otimes_{j \in \mathbf{Z}}Tr_{j})(\sigma_{x}^{0})$, where
$Tr_{j}$ denotes the normalized trace. This exemplifies the way how the limit $t \to \infty$ may exist on a state on the 
quasi-local algebra, but, for the complete argument, we must refer to the original paper by Radin \cite{Ra}. 

At the same time, we see from the above how non-recurrent behavior may arise when interactions of infinite range (which are physically
most realistic, finite range being just approximations) are taken into account.

As in \cite{Ra}, we consider the following special case in (4.1): 
$$
\epsilon(|j|) = \xi^{-|j|} \mbox{ with } \xi > 1 \mbox{ and } j \ne 0
\eqno{(4.8)}
$$
(the so-called exponential model $E_{\xi}$ in the notation of \cite{Ra}: we take for $\xi$ a number which is not transcendental, e.g., an integer). Such interactions satisfy (2.17) and thus the hypothesis of proposition 2.2. We have (\cite{Ra}, Prop. 5 and corollary page 2953):

\textbf{Proposition 4.1} With (4.1) and (4.7), let $\rho$ be any state such that
$$
\rho(\sigma^{A})=0 \forall A \mbox{ such that } A_{3} \ne \phi
\eqno{(4.9)}
$$
Then
$$
\lim_{t \to \infty} \rho_{t}(A) \equiv \bar{\rho}(A)= (\otimes_{j \in \mathbf{Z}}Tr_{j})(A)
\eqno{(4.10)}
$$
where $Tr_{j}$ denotes the normalized trace, for any $A \in \cup_{\Lambda} {\cal A}_{\Lambda}$, the local algebra. Thus (4.10) generalizes to the quasi-local algebra ${\cal A}$.

Above, in (4.9), for each triple $A=(A_{1},A_{2},A_{3})$, where the $A_{i}$ are pairwise disjoint, $\sigma^{A}$ is defined as $\prod_{i \in A_{1}}\sigma_{x}^{i}\prod_{j\in A_{2}} \sigma_{y}^{j}\prod_{k\in A_{3}}\sigma_{z}^{k}$, where $\prod_{l\in \phi} B_{l}$ is defined to be the identity $\mathbf{1}$.

\textbf{Corollary 4.2} Choose 
$$
\rho_{0}= \otimes_{j\in \mathbf{Z}} P_{j}^{+,x} \mbox{ where }\sigma_{x}^{j} P_{j}^{+,x} = P_{j}^{+,x}
\eqno{(4.11)}
$$
Then theorem 3.1 holds, with
$$ 
s(\rho_{0})= 0 
\eqno{(4.12.1)}
$$
and
$$
s(\bar{\rho}) = \log 2
\eqno{(4.12.2)}
$$

\textbf{Proof} 

The finite volume version of (2.8) satisfies, by (4.11), $s(\rho_{0}) = (-\lambda \log \lambda)(\lambda=0) = 0$ and, by (4.10) and (2.8), $s(\bar{\rho}) = (-\sum_{i=1}^{2}(\lambda_{i} \log \lambda_{i})(\lambda_{i}=1/2)= \log 2 $. q.e.d.

\textbf{Remark 4.1} 

Note that the above values of the mean entropy correspond to the two extreme values in (2.6) (with $D=2$). Of course, $P^{-,x}, P^{\pm,y}$ are all suitable in corollary 4.2. 

\textbf{Remark 4.2} 

Corollary 4.2 illustrates the important fact that the inequality in (3.2) may be strict.

\textbf{Remark 4.3} 

(3.1) is analogous to the mixing condition (1.5) in the classical case, and we therefore view the example(s) in corollary 4.2 as physically satisfactory.

\textbf{Remark 4.4} 

Temperature does not appear explicitly in corollary 4.2. In general, $\bar{\rho}$, given by (3.1), may be a non-equilibrium stationary state (NESS). This will be the case at $T=0$, since the ground state associated to (4.1) is the ferromagnetic ground state $\otimes_{j\in \mathbf{Z}} P_{j}^{\pm,z}$ (if $\nu = 1$). 
The last remark brings us to the first step (preparation) of the state in definition 1.1.

\section{The time-arrow problem}

In this section we show that the time-arrow problem is a consequence of the preparation of the state in definition 1.1. If we define
$$
\bar{\rho}^{'} \equiv \lim_{t \to -\infty} {\rho}_{t}
\eqno{(5.1)}
$$
and ignore, as before, the preparation of the state $\rho$ in definition 1.1, we obtain
$$
s(\rho_{0}) \le s(\bar{\rho}^{'})
\eqno{(5.2)}
$$
(5.2) expresses the \textbf{time-arrow} problem: the mean entropy increases, starting from a given state, in any of the two time directions. (5.2) is an immediate consequence of time-reversal symmetry $t \to -t$, which in the present case derives just from the self-adjointness of finite volume Hamiltonians in (2.17), and the property $\overline{\rho_{0,\Lambda}(A^{\dag})}= \rho_{0,\Lambda}(A)$, where the bar denotes complex conjugation, and $\dag$ the adjoint operation.

According to definition 1.1, the system is closed from $t=0$ to $t = \infty$, but not from $t=-r$ to $t=0$, where it is subject to external conditions, but is still thermally isolated. The work $W$ done by the time-dependent external forces on the system satisfies the, under assumption (1.7), 
$$
W \ge 0
\eqno{(5.3)}
$$
by the Kelvin-Planck statement of the second law. We shall assume that
$$
0 < W=U < \infty
\eqno{(5.4)}
$$
where $U$ is the energy imparted to the system. Under assumption (5.4), the case $r=0$ in definition 1.1, i.e., an instantaneous preparation of the state at $t=0$, a ''$\delta(t)$'' pulse, is excluded. This obvious physical requirement has a far-reaching consequence:

\textbf{Time arrow theorem} Under an adiabatic transformation (definition 1.1 with $r \ne 0$), there is a breakdown of time-reversal symmetry (5.2) and therefore, in general, a time-arrow exists.

It is not easy to provide concrete tractable models of the preparation of the state, according to definition 1.1. For the models considered in section 4, an adequate model satisfying (1.7), corresponding to (4.11) as initial state, is given by the finite volume Hamiltonian
$$
H_{\Lambda}(t) = H_{\Lambda} + f(t) \pi/2 \sum_{j\in \Lambda} \sigma_{y}^{j} \mbox{ where } f(t)=0 \mbox{ if } t \le -r \mbox{ or } t \ge 0
\eqno{(5.5)}
$$
where $H_{\Lambda}$ is given by (5.1) and
$$
f(t) = g(t+r/2)
\eqno{(5.6)}
$$
Above, $g$ denotes a smooth approximation to the delta function. We take for the initial state at $t=-r$ the ferromagnetic ground state 
$$
\rho_{g}= \otimes_{j\in \mathbf{Z}} P_{j}^{-,z}
\eqno{(5.7)}
$$
We take further the limit of a ''$\delta$'' pulse, acting on the finite-volume approximation to (5.7): let $\Omega_{\Lambda}$ denote the corresponding vector. The result for the evolution of this vector is
$$
\exp(iH_{\Lambda}r/2)\exp(i\frac{\pi}{2} \sum_{j\in \Lambda} \sigma_{y}^{j})\exp(iE_{0,\Lambda}r/2)\Omega_{\Lambda}
\eqno{(5.8)}
$$
where $E_{0,\Lambda}$ denotes the ground state energy of $H_{\Lambda}$. The last phase disappears upon construction of the finite-volume approximation to the state. In this way, we obtain the models of section 4, but with time translated by the quantity $r/2$. If $r$ is taken much smaller that the characteristic relaxation times of the system, the results obtained for the decay for the $x$- component of the magnetization are arbitrarily close to those in \cite{Ra}. These conditions correspond to the case of the experiment reported in (\cite{LowNo}, pg. 57), in which, while the $\pi/2$ rf pulse is applied to the nuclei in the calcium-fluorine crystal, the macroscopic magnetization is not decaying but has a constant magnitude. It should be emphasized that the ''sudden'' interaction we considered was done solely for technical reasons, as a caricature of the adequate model (5.5), which complies with (5.4).      

Our last section recalls the role of temperature, which, for equilibrium states, is well-known to be related to the second law through the important concept of passivity \cite{PuWo}. We comment on possible generalizations, having in mind ''slight'' deviations of equilibrium (temperature) states.

\section{Possible generalizations: slight deviations of thermal equilibrium states and quantum continuous systems}

Concerning thermal equilibrium states $\rho^{eq}$, it is natural to inquire whether the second law holds for locally perturbed states, in the folium of $\rho^{eq}$ (\cite{BRo1}, \cite{BRo2}, \cite{Sewell1}), i.e., such that
$$
\rho_{A}(B) = \frac{\rho^{eq}(A^{*}B A)}{\rho^{eq}(A^{*}A)} \mbox{ with } \rho^{eq}(A^{*}A) \ne 0 \mbox{ and } A,B \in {\cal A} 
\eqno{(6.1)}
$$
By lemma 3.5 of \cite{NTh4}, for any primary or factor state (see \cite{BRo2} or \cite{Sewell1}) $\rho^{eq}$, invariant under a time translation automorphism $\alpha_{t}$ of the quasilocal algebra ${\cal A}$, the following three properties are equivalent:
$$
\lim_{t\to \pm \infty} \rho_{A}(\alpha_{t}(B)) = \rho^{eq}(B) \forall A,B \in {\cal A} 
\eqno{(6.2)}
$$
$$
\lim_{t\to \pm \infty} \rho^{eq}(A \alpha_{t}(B)C) = \rho^{eq}(AC) \rho^{eq}(B) \forall A,B,C \in {\cal A} \mbox{ (generalized mixing) }
\eqno{(6.3)}
$$
$$
\lim_{t \to \pm \infty} \rho^{eq} (A[\alpha_{t}(B),C]D) = 0 \mbox{ (weak asymptotic abelianness) }
\eqno{(6.4)}
$$
The very interesting remark was made by Narnhofer and Thirring \cite{NTh4} that the opposite requirements of the system being classical for large times (weak asymptotic abelianness (6.4)) and being completely quantal, i.e., a factor state with center ${\cal Z}_{\rho}=\{z \mathbf{1}\}$ (see again \cite{BRo2}\cite{Sewell1}) constrain it to the extent that all observables have to approach their equilibrium values (generalized mixing), in sharp contrast to the classical case. The authors conjecture in \cite{NTh4}, pg. 2949, that Galilei or Poincar\`{e} invariance may exclude the fact that some finite parts of the system are somewhere ''locked'', precluding weak asymptotic abelianness, as might occur for lattice systems (the fact that this occurs was shown by Radin in proposition 3 of \cite{Ra}). In \cite{NTh4} the authors proved that a class of Galilei invariant fermion systems does satisfy (6.4), and in \cite{JNW} the same fact was proved for a class of two-dimensional thermal quantum field theories.

It would be thus of special interest to prove that a theorem such as theorem 3.1 holds for local perturbations of primary thermal equilibrium states of continuous quantum fermion systems, such as those treated in \cite{LanRo}, but with the additional requirement of Galilei invariance, for which (6.2) is expected to hold as a consequence of weak asymptotic abelianness.

Finally, it is to be noted that the state at the final step of the preparation process at $t=0$ was not a local perturbation of the equilibrium state at the beginning $t=-r$ of the process, in the cases treated in section 4: for instance, the states (4.11) and (5.7) are disjoint.

\section{Conclusion}

We proposed a dynamical proof of the second law, which depends crucially on the upper semicontinuity of the specific entropy. Basic to the approach are the universal form (Gibbs-von Neumann) for the entropy, and viewing an adiabatic process as the evolution through a Hamiltonian dynamics (classical or quantum)  of a (non-invariant) state, which tends to an invariant state, with, in general, higher mean entropy. In general, this latter state is a non-equilibrium invariant state (NESS), and the initial state is not a ''gentle'' perturbation of the final state.

The present approach seems adequate to explain irreversibility (remark 3.1) and the time-arrow (section 5), without appealing to randomness or conditions of the Stosszahlansatz type. Although preparation is essential to explain the time-arrow, the assumptions of adiabatic transformation (definition 1.1) do not require the presence of an observer, because the required energy in (5.5) may be supplied by the natural environment. An example is given by hydrogen in interstellar space \cite{Carruthers}: the energy necessary to produce the resonance states of atomic and molecular hydrogen comes, of course, from the radiation emitted by the stars. Another example is natural radioactivity, in connection with alpha decay. In both cases the initial time plays no role in the determination of the values of the physical parameters (energy levels, half-lives), because the Lorentzian approximation is excellent over many half-lives (see \cite{Wre}). An extension of the model studied in  \cite{Wre}, to the ''recombination epoch'', for an initial state close to a plasma of free electrons and protons, should yield irreversibility in the sense of theorem 3.1, with the r.h.s. of (3.2) given by the sum of the specific entropies  of free thermal $3K$ photons and the specific entropy of free hydrogen atoms in their ground state, the latter equal to zero. As in the case of the final state in corollary 4.2, which is metastable due to its being stationary, the thermal photons of the cosmic microwave background (CMB) are metastable, because they interact with the hydrogen atoms in their ground state only through the tail of the Planck distribution (see the discussion in \cite{Harrison}, pg. 347). Unfortunately, we are far from a rigorous discussion of such a model, but should like to point out the essential importance of metastable states in the Universe.

The independence of the role of the observer has been emphasized by Haag \cite{Haag} and the ET- H approach (\cite{Fro} and references given there). Actually, the observer only measures the ''future'' by definition, if the ''future'' refers to the psychological sensation that time passes, but this time should coincide with the objective physical future, as emphasized by Haag \cite{Haag}. Given that human beings are also unstable systems, which require a ''preparation'', this assumption seems to fit in the present approach.

The ''coarse-graining'' which is necessary for irreversible behavior is of both kinds: in time and in space, the latter due to the necessity of considering the specific entropy, rather than the entropy, both are related, and no further sources of coarse graining are required. This point is our main qualitative contribution in theorem 3.1. We believe that the most important issue lies in the complexities of both the classical (\cite{LP}, \cite{Sinai1}) and the quantum evolutions \cite{NT}, \cite{Em2}: quantum K systems, studied in the latter two references, were proved there to display the ''memoryless'' behavior known in generic classical dynamics.

\section{Acknowledgements} We should like to thank Lawrence Landau, Heide Narnhofer and Derek W. Robinson for their remarks in a fruitful correspondence. The remarks of professors Oliver Penrose and David Ruelle in correspondence are also gratefully acknowledged. 

In a previous version we overlooked the fact that the density matrices $(\rho_{t})_{\Lambda}$ and $\rho_{\Lambda,t}$, the latter given by (2.10), are not the same, unless there are no interactions. This was pointed out to us by Lawrence Landau, as well as by the reviewer. We are very grateful to the reviewer, who generously provided us with propositions 2.1 and 2.2, which prove that this difference indeed does not affect the specific entropy. He should be considered a co-author of this paper.

\end{document}